\documentclass[prd,reprint,showpacs,superscriptaddress,preprintnumbers]{revtex4-1}

\usepackage{amsmath}
\usepackage{amstext}
\usepackage{amsfonts}
\usepackage{amssymb}
\usepackage{graphicx}
\usepackage{hyperref}
\usepackage{pstricks}
\usepackage{pst-plot}
\usepackage{pstricks-add}

\newcommand{\CP}{\(\mathrm{CP}\)}
\newcommand{\hc}{\text{h.}\,\text{c.}}
\newcommand{\lt}{\left}
\newcommand{\rt}{\right}
\newcommand{\gev}{\,\mbox{GeV}}
\newcommand{\tev}{\,\mbox{TeV}}

\newcommand{\fig}[1]{Fig.~\ref{#1}}

\newcommand{\eq}[1]{(\ref{#1})}

\newcommand{\bra}[1]{\ensuremath{\langle #1 |}}
\newcommand{\ket}[1]{\ensuremath{| #1 \rangle }}
\newcommand{\ov}{\overline}
\newcommand{\Bbar}{\,\overline{\!B}}
\newcommand{\bbs}{\ensuremath{B_s\!-\!\Bbar{}_s\,}}
\newcommand{\bbms}{\bbs\ mixing}

\begin{document}
\title{Vacuum stability of the effective Higgs potential in the \\
  Minimal Supersymmetric Standard Model}

\author{Markus Bobrowski} \email{markus.bobrowski@kit.edu}
\affiliation{Institut f\"{u}r Theoretische Teilchenphysik, Karlsruhe Institute
  of Technology, Engesserstra\ss{}e 7, 76128 Karlsruhe, Germany}

\author{Guillaume Chalons} \email{chalons@lpsc.in2p3.fr}
\affiliation{Institut f\"{u}r Theoretische Teilchenphysik, Karlsruhe Institute
  of Technology, Engesserstra\ss{}e 7, 76128 Karlsruhe, Germany}
\affiliation{{Laboratoire de Physique Subatomique et de Cosmologie, Universit\'e
Grenoble-Alpes, CNRS/IN2P3, 53 Avenue des Martyrs, F-38026 Grenoble, France}}

\author{Wolfgang G.~Hollik} \email{wolfgang.hollik@kit.edu}
\affiliation{Institut f\"{u}r Theoretische Teilchenphysik, Karlsruhe Institute
  of Technology, Engesserstra\ss{}e 7, 76128 Karlsruhe, Germany}

\author{Ulrich Nierste} \email{ulrich.nierste@kit.edu}
\affiliation{Institut f\"{u}r Theoretische Teilchenphysik, Karlsruhe Institute
  of Technology, Engesserstra\ss{}e 7, 76128 Karlsruhe, Germany}

\begin{abstract}
  The parameters of the Higgs potential of the Minimal Supersymmetric
  Standard Model (MSSM) receive large radiative corrections which lift
  the mass of the lightest Higgs boson to the measured value of 126\gev.
  Depending on the MSSM parameters, these radiative corrections may also
  lead to the situation that the local minimum corresponding to the
  electroweak vacuum state is not the global minimum of the Higgs
  potential. We analyze the stability of the vacuum for the case of
  heavy squark masses as favored by current LHC data. To this end we
  first consider an effective Lagrangian obtained by integrating out the
  heavy squarks and then study the MSSM one-loop effective potential
  $V_{\rm eff}$, which comprises all higher-dimensional Higgs couplings
  of the effective Lagrangian. We find that only the second method gives
  correct results and argue that the criterion of vacuum stability
  should be included in phenomenological analyses of the allowed MSSM
  parameter space.  Discussing the cases of squark masses of 1 and 2
  \tev\ we show that the criterion of vacuum stability excludes a
  portion of the MSSM parameter space in which $|\mu \tan\beta|$ and $|A_\text{t}|$
  are large.
\end{abstract}

\preprint{TTP14-018}
\preprint{LPSC-14-130}

\maketitle

\section{Introduction}
\begin{figure*}[t]
\includegraphics{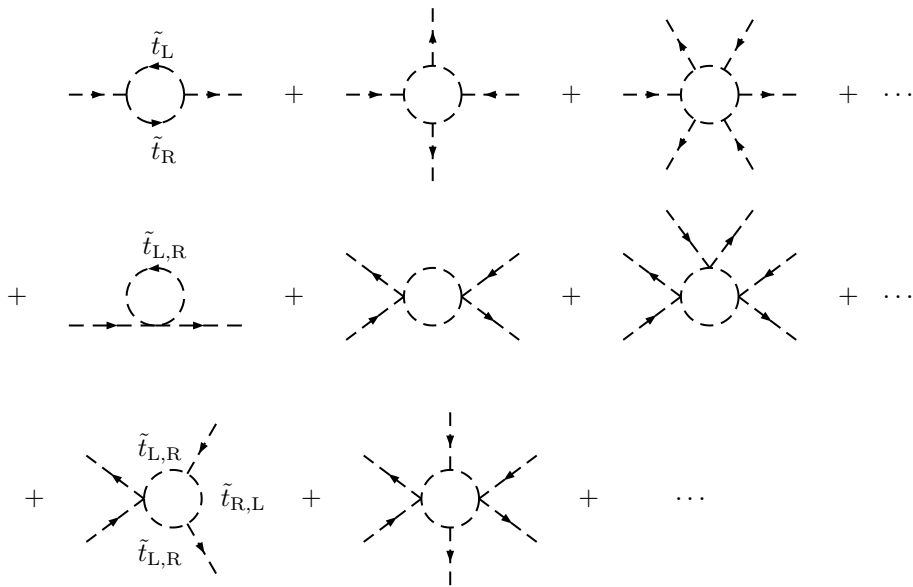}
  \caption{The 1-loop contribution to the effective potential as the sum
    of all one-particle irreducible diagrams with zero external
    momenta.}
  \label{fig:diagrams}
\end{figure*}

The Higgs sector of the Minimal Supersymmetric Standard Model (MSSM)
comprises two Higgs doublets $H_\text{u}$ and $H_\text{d}$ with
tree-level Yukawa couplings to up-type and down-type fermions, respectively.
Their self-interaction is described by a special version of the
Higgs potential of a two-Higgs-doublet model (2HDM)
\cite{HaberKane, Gunion:2002zf}:
\begin{equation}\label{eq:V2HDM}
\begin{aligned}
  V &=m_{11}^{2}\;H_\text{d}^{\dagger}H_\text{d} +
  m_{22}^{2}\;H_\text{u}^{\dagger}H_\text{u} + \left(
    m_{12}^{2} \; H_\text{u}\cdot H_\text{d}+ \hc \right) \hfill\\[1ex]
  & +\frac{\lambda_{1}}{2}\big(H_\text{d}^{\dagger}H_\text{d}\big)^{2}+
  \frac{\lambda_{2}}
  {2}\big(H_\text{u}^{\dagger}H_\text{u}\big)^{2}+\lambda_{3}\big(H_\text{u}^{\dagger}
  H_\text{u}\big)\big(H_\text{d}
  ^{\dagger}H_\text{d}\big) \hfill\\[1ex]
  & +\lambda_{4}\big(H_\text{u}^{\dagger}H_\text{d}\big)
  \big(H_\text{d}^{\dagger}
  H_\text{u}\big)+\bigg( \frac{\lambda_{5}}{2}\big( H_\text{u}\cdot H_\text{d}
    \big)^{2}\hfill\\[1ex]
 & -\lambda_{6}\big(H_\text{d}^{\dagger}H_\text{d}\big)\big(
    H_\text{u}\cdot H_\text{d} \big)  -\lambda_{7}\big(H_\text{u}^{\dagger}H_\text{u}\big)\big(
    H_\text{u}\cdot H_\text{d} \big) + \hc \bigg).
\end{aligned}
\end{equation}
The neutral components of $H_\text{u,d}$ acquire vacuum expectation
values (vevs) $v_\text{u,d}/\sqrt{2}$ satisfying $\sqrt{v_\text{u}^2 +
  v_\text{d}^2}=v\simeq 246\gev$.  In the MSSM the tree-level values for
the self-couplings $\lambda_{1\ldots 4}$ are fixed in terms of small gauge
couplings and those of $\lambda_{5\ldots 7}$ vanish altogether. As a
consequence, the mass of the lightest Higgs boson $h^0$ cannot exceed
the $Z$-boson mass at tree level. Radiative corrections can lift $m_{h^0}$
well above $m_Z$ \cite{Haber:1990aw} and must indeed be large, if the
discovered Higgs boson with a mass of 125$\gev$
\cite{Chatrchyan:2012ufa,Aad:2012tfa} is identified with $h^0$. The
largest radiative corrections to $m_{h^0}$ involve the top Yukawa
coupling $Y_\text{t}$ and stem from loop diagrams with stops or tops.
Diagrammatic two-loop
\cite{Hempfling:1993qq,Heinemeyer:1998jw,Heinemeyer:1998kz,Heinemeyer:1998np} and three-loop
\cite{Harlander:2008ju,Kant:2010tf} corrections to $m_{h^0}$ are
implemented in the public computer programs {\tt FeynHiggs}
\cite{Heinemeyer:1998np,Heinemeyer:1998yj,Degrassi:2002fi,Frank:2006yh,Hahn:2013ria}
and {\tt H3m} \cite{Kant:2010tf}, respectively.
No stops at the LHC have been found, suggesting that the masses $m_{\tilde t_{1,2}}$
of the two stop eigenstates are well above the electroweak scale
$v$. Heavy stops require large values for the bilinear
supersymmetry-breaking terms $m_{\tilde t_{\rm L,R}}^2$, which are the
diagonal elements of the stop mass matrix.  In the limit $m_{\tilde
  t_{\rm L,R}}^2\gg v$ one can integrate out the heavy stops to find an
effective 2HDM Lagrangian ${\cal L}_{\rm 2HDM}\supset -V$, which encodes
the stop effects in terms of effective parameters $m_{ij}^{2}$ and
$\lambda_i$.  To derive $V$ one must calculate diagrams with two or four
external Higgs lines and a stop loop. The $\lambda_i$ receive shifts
proportional to $Y_\text{t}^4$ which are crucial to lift $m_{h^0}$ to the
measured value. If $\tan\beta =v_\text{u}/v_\text{d}$ (or the trilinear supersymmetry-breaking term
$A_\text{b}$) is large, also sbottom loops must be considered. An exhaustive
analysis, matching the MSSM with heavy superpartners onto a 2HDM at the
full one-loop level can be found in \cite{Gorbahn:2009pp}.
Denoting the masses of the top and bottom squarks generically with
$M_{\tilde q}$, the Higgs masses and couplings calculated from the
effective 2HDM reproduce the results of the diagrammatic calculation as
an expansion in $1/M_{\tilde q}^2$. The accuracy of this expansion can
be improved by adding terms of higher dimension to \eq{eq:V2HDM}
obtained from loop diagrams with more external legs as shown in
\fig{fig:diagrams}. Effective Lagrangians permit the resummation of
large logarithms $\ln (M_{\tilde q}/v)$ to all orders in perturbation
theory by solving the renormalization-group (RG) equations for the
parameters. Note that the top quark is not integrated out, ${\cal
  L}_{\rm 2HDM}$ contains the full field content of the 2HDM and
\emph{e.g.}\ top-loop contributions to the Higgs mass matrix are calculated
from ${\cal L}_{\rm 2HDM}$.

The effective 2HDM lagrangian reproduces the low-energy ($E\ll M_{\tilde
  q}$) phenomenology of the MSSM for the case
$v,m_{h^0},m_{A^0},m_{H^0},m_{H^\pm}\ll M_{\tilde q}$. The Yukawa sector
of ${\cal L}_{\rm 2HDM}$ has been widely studied
\cite{Banks:1987iu,Hall:1993gn,Carena:1994bv,Blazek:1995nv,Hamzaoui:1998nu,
  cgnw,cgnw2,bcrs,Isidori:2001fv,Buras:2002wq,Dedes:2002er,brs,cn,
  Hofer:2009xb,Gorbahn:2009pp,Crivellin:2010er,Crivellin:2011jt,
  Crivellin:2012zz}, while little attention has been devoted to the
Higgs self-interaction in $V$. Instead, effective-Lagrangian studies
(typically addressing calculations of $m_{h^0}$) have used a Higgs
potential with a single Higgs doublet, describing instead the hierarchy
$v,m_{h^0} \ll m_{A^0},m_{H^0},m_{H^\pm},M_{\tilde q}$,
\emph{i.e.}\ integrating out the heavy non-standard Higgs fields at the same
scale as the heavy superpartners
\cite{Espinosa:1991fc,Carena:2000dp}. In \cite{Hahn:2013ria} the
diagrammatic two-loop result for $m_{h^0}$ of \cite{Degrassi:2002fi} is complemented
with the leading and next-to-leading logarithms $\ln (M_{\tilde q}/v)$
of higher orders found from the RG analysis of the single-Higgs-doublet
Lagrangian in \cite{Espinosa:1991fc, Arason:1991ic}. The corresponding
result is implemented in the current version \texttt{2.10.0} of
\texttt{FeynHiggs}.
Other public computer codes incorporating two-loop accuracy
  for the Higgs boson mass are \texttt{Softsusy} \cite{Allanach:2001kg},
  \texttt{SuSpect} \cite{Djouadi:2002ze} and \texttt{SPheno} \cite{Porod:2003um}.

Depending on the values of the $\lambda_i$ the 2HDM potential in
  \eq{eq:V2HDM} may be unbounded from below (UFB) or may develop an
  unwanted global minimum rendering ``our'' vacuum state with
  $v=246\gev$ unstable. The parameter ranges complying with vacuum
  stability have been identified in \cite{Gunion:2002zf,Barroso:2013awa}
  and the corresponding constraints on $m_{ij}^2$ and $\lambda_i$ are
  routinely included in phenomenological analyses of 2HDM (see
  \emph{e.g.}~{\cite{Eriksson:2010zzb,Krawczyk:2013gia,Eberhardt:2013uba,Baglio:2014nea}}).
  These vacuum stability
  constraints can also be imposed on the effective ${\cal L}_{\rm 2HDM}$
  obtained from the MSSM by integrating out heavy squarks. In this paper
  we show that there are indeed ranges for the MSSM parameters for which
  $V$ in \eq{eq:V2HDM} is unbounded from below. However, when $V$ drops
  below its local minimum with $v=246\gev$ the Higgs fields are so large
  that the higher-dimensional corrections to $V$ depicted in
  \fig{fig:diagrams} become important. All these contributions can be
  resummed and constitute a piece of the effective Coleman-Weinberg
  potential \cite{Coleman:1973jx}. In \cite{Casas:1995pd,Casas:1996de,Casas:1997ze}
the multiple minima of the full tree-level scalar potential have been surveyed in great
detail and strong constraints on their existence were derived. Analyses allowing the
electroweak vacuum to be metastable, with a lifetime exceeding the age of the universe,
have been performed in
\cite{Kusenko:1996jn,Camargo-Molina:2013pka,Chowdhury:2013dka,Blinov:2013fta,Camargo-Molina:2013qva,Camargo-Molina:2014pwa}.
While the effective Higgs potential
  for the MSSM has been widely studied  \cite{Okada:1990vk,
      Ellis:1990nz,Barbieri:1990ja,Ellis:1991zd,Brignole:1991pq,Carena:2000dp,
      Espinosa:1991fc,Haber:1993an,Carena:1995bx,Carena:1995wu,
      Martin:2002i} with focus on Higgs masses,
  the criterion of vacuum
  stability of the loop-corrected Higgs potential has previously not been applied to
constrain the MSSM
  parameter space. The studies performed in
\cite{CamargoMolina:2012hv,Camargo-Molina:2013qva,Camargo-Molina:2014pwa} used the {\tt
Vevacious} code to exploit the vacuum stability constraint on the parameter space. This
code makes use of the effective Coleman-Weinberg potential but only in the vicinity of
all the vacua found by minimization of the \textit{tree-level} scalar potential. However,
this procedure does not guarantee to find minima induced purely by radiative effects
\cite{Camargo-Molina:2013qva}, which is precisely the topic we analyze in this paper.

 This paper is organized as follows: In Sec.~\ref{sec:effl}
we re\-derive the effective potential of the MSSM and discuss
some of its properties. In Sec.~\ref{sec:ph} we illustrate the main
result of this paper, a novel constraint on the MSSM parameter space
from the requirement of vacuum stability. Finally we conclude.

\section{Effective Lagrangian and effective potential}
\label{sec:effl}
Once we integrate out the heavy top and bottom squarks we find
the \emph{effective Lagrangian}
\begin{eqnarray}
{\cal L}_{\rm 2HDM}^{\text{{Higgs}}} &=& {\cal L}_{\rm kin} -V + {\cal L}_{\rm
der,D\geq 6} -
 V_{D\geq 6}
\end{eqnarray}
with the kinetic term ${\cal L}_{\rm kin}$, the Higgs potential $V$ of
\eq{eq:V2HDM}, and the contribution from higher-dimensional operators
$ {\cal L}_{\rm der,D\geq 6} - V_{D\geq 6}$. The former term
$ {\cal L}_{\rm der,D\geq 6}$ contains operators of dimension 6 and
higher with at least two derivatives acting on the Higgs fields. One
effect of $ {\cal L}_{\rm der,D\geq 6}$ are contributions suppressed
by one or more powers of $1/M_{\tilde q}^2$
to the field renormalization constants of the physical Higgs fields.
(These renormalization constants can be matrices,
permitting kinetic mixing of different Higgs fields.) This effect
matters for the expression of the doublet components in terms of
physical fields, but is of no relevance for the discussion of global
properties of the Higgs potential in this paper. Other ingredients of
${\cal L}_{\rm der,D\geq 6}$ are derivative couplings and couplings to gauge
fields, which are also irrelevant for our analysis. The Higgs potential
$V+V_{D\geq 6}$ contains the usual bilinear and quadrilinear
tree-level contributions and the loop contributions depicted
in \fig{fig:diagrams}. If the effective Lagrangian is used to calculate
Higgs masses and mixing angles, the series of higher-dimensional terms
in ${\cal L}_{\rm 2HDM}$ will give corrections which quickly decrease
with powers of $1/M_{\tilde q}^2$.

Our purpose, however, is to study global properties of $V+V_{D\geq 6}$
and $V_{D\geq 6}$ can be sizable in the range of large Higgs field
amplitudes. It is well-known how to resum the contributions with
$D=6,8,10\ldots$, the result is the squark contribution to the
\emph{effective potential}. The concept of the effective potential does
not require a mass hierarchy between the particles running in the loop
and the external Higgs bosons and indeed the original application of
Coleman and Weinberg \cite{Coleman:1973jx} involves a massless field in
the loop. The focus of \cite{Coleman:1973jx} is the generation of a
small dynamical Higgs mass in a theory with zero tree-level mass through
spontaneous symmetry breaking induced by quantum effects, which are
subsumed in the effective potential. Instead the scope of our paper is the
destabilization of the tree-level MSSM Higgs potential by very heavy
particles (top and bottom squarks). Still, as in the original paper we
use the effective potential to ``survey all possible vacua
simultaneously'' \cite{Coleman:1973jx}.

In Sec.~\ref{sec:el} we calculate the higher-dimensional couplings of
neutral Higgs bosons in $V_{D\geq 6}$.  In Sec.~\ref{sec:ep} we
summarize some of the conceptual aspects of the effective potential
and show that the one-loop effective potential of the
  MSSM \cite{Brignole:1991pq} indeed reproduces the couplings derived in
  Sec.~\ref{sec:el} correctly.

\subsection{Effective 2HDM Lagrangian}\label{sec:el}

\begin{figure*}[t]
\includegraphics{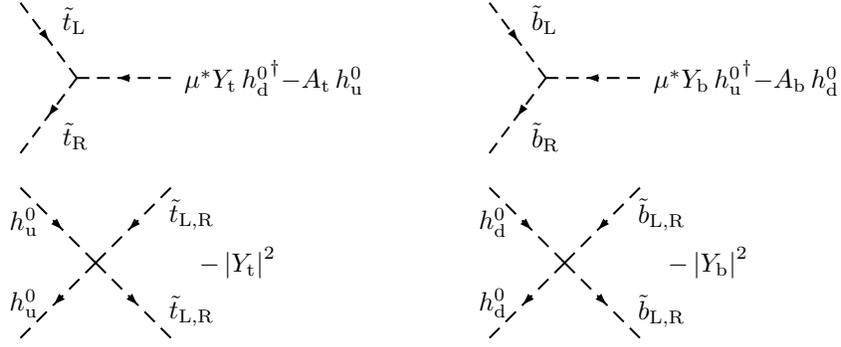}
  \caption{Couplings of neutral Higgs fields to squarks in the MSSM.
   $\mu$ is the higgsino mass parameter and $Y_{\text{t,b}}$ and
   $A_{\text{t,b}}$ are the Yukawa coupling and trilinear SUSY-breaking term,
   respectively, of top or bottom (s)quarks.}
  \label{fig:vertex}
\end{figure*}

The two $\text{SU}(2)$ doublet Higgs fields of the MSSM are
\begin{equation}\label{eq:HiggsDoublets}
 H_\text{u} = \left(
  \begin{array}{c}
    h_\text{u}^+ \\[.5ex] h_\text{u}^0\\
  \end{array}\right), \qquad
 H_\text{d} = \left(
  \begin{array}{c}
    h_\text{d}^0 \\[.5ex]  - h_\text{d}^-\\
  \end{array}\right)
\end{equation}
with hypercharges $+1/2$ and $-1/2$, respectively, and vevs
$\langle h_\text{u}^0\rangle = v_\text{u}/\sqrt{2}$ and $\langle
h_\text{d}^0\rangle = v_\text{d}/\sqrt{2}$. As usual we define their
  ratio as $\tan\beta =v_\text{u}/v_\text{d}$. The most general
  renormalizable Higgs potential $V$ of an arbitrary 2HDM
\cite{HaberKane, Gunion:2002zf} is given in
  \eq{eq:V2HDM} above, where $a\cdot b = a^\text{T}\,\epsilon\, b$ and
\(\epsilon\) denotes the totally antisymmetric tensor with
\(\epsilon_{12} = +1\).
At tree-level, $V$ is unambiguously determined by \(F\)- and \(D\)-terms and the
soft supersymmetry breaking Lagrangian:
\begin{equation}
\begin{aligned}
m_{11}^{2\,^{\rm tree}} & = \left\vert \mu\right\vert
^{2}+m_{H_\text{d}}^{2},
& \lambda_{1,2}^{\rm tree}  &=-\lambda_{3}^{\rm tree}=\frac{g^{2}+g^{\prime2}}{4} , \\[.5ex]
m_{22}^{2\,^{\rm tree}} & = \left\vert \mu\right\vert
^{2}+m_{H_\text{u}}^{2},
& \lambda_{4}^{\rm tree}  & = \frac{g^{2}}{2}, \\[.5ex]
m_{12}^{2\,^{\rm tree}} & = B_\mu ,
& \lambda_{5}^{\rm tree}  & =\lambda_{6}^{\rm tree}=\lambda_{7}^{\rm tree}=0.
\end{aligned}\label{eq:MSSM-rep}
\end{equation}
{With the minimization conditions one can eliminate
$m_{11}^2$ and  $m_{22}^2$ in terms of $v$ and $\beta$. At tree-level,
these relations read
\begin{equation}\label{eq:m11vev}
\begin{aligned}
m_{11}^{2\,^{\rm tree}} &= m_{12}^{2\,^{\rm tree}} \tan\beta -
              \frac{v^2}{2} \cos(2\beta) \lambda_1^{\rm tree} ,\\[.5ex]
m_{22}^{2\,^{\rm tree}} &= m_{12}^{2\,^{\rm tree}} \cot \beta +
              \frac{v^2}{2} \cos(2\beta) \lambda_1^{\rm tree}.
\end{aligned}
\end{equation}
One further has the relation
$2 m_{12}^{2\,{\rm tree}}=m_A^{2\,{\rm tree}} \sin(2\beta)$, where $m_A^{\rm tree}$
  is the tree approximation to the mass of the pseudoscalar Higgs boson
  $A^0$. This relation and those in \eq{eq:m11vev} change once
radiative corrections are included, \emph{e.g.}\ the formulae are affected
by loop corrections to $\lambda_{1\ldots3,\,5\ldots7}$ (see eqs.~(23)--(27) of
\cite{Gorbahn:2009pp}) and the parameters of $V_{D\geq 6}$.}
For the following discussion it is useful to write
\begin{equation}
\lt.  \phantom{\frac12} \lt(V+ V_{D\geq 6}\rt)
\rt|_{h_{u,d}^\pm\to 0} \, =\, V_0 + V_1,
\end{equation}
where $V_0$ and $V_1$ denote the tree and one-loop contributions,
respectively, and the subscript on the LHS means that the charged
Higgs fields are set to zero.  $V_0$
equals $V$ with the parameters in \eq{eq:MSSM-rep}, while $V_1$ is
obtained from the sum of one-loop diagrams in Fig.~\ref{fig:diagrams}.
Neglecting loops with small gauge couplings (which are kept in
  the tree-level terms) and retaining only the stop loop for the moment
the result has the schematic form
\begin{equation}\label{eq:PotentialFromGreensFunctions}
    V_1 = {-}\sum\limits_{k=0}^\infty \sum_{n=0}^{{\infty}} a_{kn}
\left(h^\dagger h\right)^{k} \left(h_\text{u}^{0\dagger} h_\text{u}^0\right)^{n}.
\end{equation}
Here $h=h_{\text{d}}^{0\dagger}{-}h^{{0}}_\text{u} A_\text{t}/(\mu^* Y_\text{t})$ is
the linear combination of neutral Higgs fields coupling to the stop
loop (see \fig{fig:vertex})
and $a_{kn}$ is calculated from one-particle irreducible one-loop
diagrams with ${2k+2n}$ legs; $k$ denotes the number of $\tilde
t_\text{R}^*$-$\tilde t_\text{L}$-$h$ vertices (and equally many $\tilde
t_\text{L}^*$-$\tilde t_\text{R}$-$h^\dagger$ vertices) and ${n}$ is
the number of $\tilde t_\text{L/R}^*$-$\tilde
t_\text{L/R}$-$h_\text{u}^{0\dagger}$-$ h_\text{u}^0$ vertices. We only considered
field configurations with $h_\text{u}^+ = h_\text{d}^- = 0$.  Relaxing this constraint might exclude
additional parts of the MSSM parameter space (corresponding to charge-breaking minima),
but according to \cite{Casas:1995pd}, such minima play no significant role for the
analysis. The sbottom contribution (relevant only for large $\tan \beta$ or large
$A_\text{b}$) adds to \eq{eq:PotentialFromGreensFunctions} an analogous term
with $h$ representing a different linear combination of
$h_\text{u}^\dagger$ and $h_\text{d}$
and $h_\text{u}^0$ replaced by
  $h_\text{d}^0$. The coefficients $a_{kn}$ read
\begin{widetext}
\begin{equation}
\begin{aligned}
\label{eq:akn}
a_{kn}=& |\mu|^{2k} |Y_\text{t}|^{2n+2k} \,
   \frac{1}{k} \sum_{j=0}^n \frac{(j+k-1)!}{j!(k-1)!}
              \frac{(n-j+k-1)!}{(n-j)!(k-1)!}\,
    I_{k+j,k+n-j} (M^2_{\tilde Q}, M^2_{\tilde{\text{t}}})
         \mbox{ for }{n,k\geq1}, \hfill \\[1ex]
a_{k0} =& |\mu Y_\text{t}|^{2k} \, \frac{1}{k} \,
 I_{k,k}(M^2_{\tilde Q}, M^2_{\tilde{\text{t}}})\mbox{ for }k\geq 1, \hfill \\[1ex]
a_{0n} =& |Y_\text{t}|^{2n}  \frac1n \hspace{-.2ex}\lt[
 I_{n,0} ({M^2_{\tilde Q}}) \hspace{-.2ex}+\hspace{-.2ex}
 I_{0,n} ({M^2_{\tilde{\text{t}}}}) \rt]
         \mbox{ for }n\geq 1.
\end{aligned}
\end{equation}
\end{widetext}
Here $I_{p,q}(M^2_{\tilde Q}, M^2_{\tilde{\text{t}}})$
is the result of the one-loop diagram
with $p$ propagators of $\tilde t_\text{L}$ and $q$ propagators of $\tilde t_\text{R}$:
\begin{equation}
\begin{aligned}
 I_{p,q}(M^2_{\tilde Q}, M^2_{\tilde{\text{t}}})
   \,=\,&  \frac{3}{16\pi^2}
    \frac{1}{(p-1)!(q-1)!} \times  \hfill \\[1ex]&
    \frac{\partial^{p-1}}{\partial (M^2_{\tilde Q})^{p-1}}
    \frac{\partial^{q-1} }{\partial (M^2_{\tilde{\text{t}}})^{q-1}}
    \frac{A_0(M^2_{\tilde Q})-A_0(M^2_{\tilde{\text{t}}})}{M^2_{\tilde
        Q}-M^2_{\tilde{\text{t}}}}   \hfill \\[.5ex] & \hspace{.5\columnwidth} \mbox{for }q,p\geq 1, \hfill \\[1.5ex]
 I_{n,0}({M^2}) \,=\,&  \frac{3}{16\pi^2}
    \frac{1}{(n-1)!} \frac{\partial^{n-1}}{\partial ({M^2})^{n-1}}
    A_0({M^2})\hfill \\[.5ex] & \hspace{.5\columnwidth} \mbox{for }n\geq 1, \hfill \\[1.5ex]
 I_{0,n}({M^2}) \,=\,&
    I_{n,0}( {M^2}).
\end{aligned}
\end{equation}
In this equation $I_{p,q}$ is expressed in terms of derivatives of the
tadpole function, which equals ${A_0(M^2)=M^2(1-\ln(M^2/Q^2))}$ when
evaluated at the scale $Q$ in the $\ov{\rm MS}/\ov{\rm DR}$ scheme.  The derivation
of $a_{k0}$ and $a_{0n}$ is straightforward, the calculation of the
combinatorial factors can be found in standard textbooks. To understand
$a_{kn}$ for the case with both non-zero $k$ and $n$, consider first a
diagram with $k\neq 0$ and $n=0$, depicted in the first row of
\fig{fig:diagrams}. There are $k! (k-1)!$ diagrams (giving identical
results for zero external momenta). After dividing off the combinatorial
factor $(k!)^2$ associated with the field monomial in
\eq{eq:PotentialFromGreensFunctions}, one verifies the factor of $1/k$
in \eq{eq:akn}. These loops with only 3-point vertices have $k$ propagators of $\tilde
t_\text{L}$ and equally many propagators of $\tilde t_\text{R}$. Starting from such a
loop we now attach $n$ four-point vertices to the diagram, \emph{i.e.}\ we pass
from the first to the third row in \fig{fig:diagrams}. The sum in
\eq{eq:akn} takes care of the possibilities to place $j$ four-point
vertices on a $\tilde t_\text{L}$ line and $n-j$ such vertices on a
$\tilde t_\text{R}$ line. There are $(j+k-1)!/(j!(k-1)!)$ possibilities for the $j$
placements on a $\tilde t_\text{L}$ line, and $(n-j+k-1)!/((n-j)!(k-1)!)$ ways
to place the remaining $n-j$ vertices. (These factors correspond to a
standard exercise of combinatorics and count the number of orderless
configurations with repetitions of $j$ balls having $k$ possible colors.) Finally
there are $(n!)^2$ ways to connect the added 4-point vertices with the
external $h_\text{u}^0$ and $h_\text{u}^{0*}$ fields, which matches the combinatorial
factor of the field monomial in \eq{eq:PotentialFromGreensFunctions}.

The calculation of $a_{kn}$
  and the resummation can be
  elegantly done  with techniques developed in the effective potential
  approach used in Sec.~\ref{sec:ep}. We nevertheless find it
  instructive to calculate $a_{kn}$ explicitly as described above and to verify
  that the effective-potential method reproduces the result correctly.

\subsection{Effective potential}
\label{sec:ep}
To resum the series in \eq{eq:PotentialFromGreensFunctions}
one defines particle masses which depend
on the Higgs fields of the theory. We need the stop
mass matrix
\begin{equation}\label{eq:fielddependentmass}
\mathcal{M}^2_{\tilde{\mathrm{t}}} = \left(
  \begin{array}{cc}
 {M_{\tilde Q}^2} + \left|Y_\text{t}\, h_\text {u}^0\right|^2 & -
 \mu^\ast Y_\text{t}\, {h_\text {d}^0}^\dagger {+} A_\text{t}\, h_\text
    {u}^0 \\[1ex] - \mu Y_\text{t}^\ast\, h_\text {d}^0 {+}
    A_\text{t}^\ast\, {h_\text {u}^0}^\dagger&
    {M_{\tilde{\text{t}}}^2} + \left|Y_\text{t}\,
    h_\text{u}^0\right|^2\\
 \end{array}   \right),
\end{equation}
where $M_{\tilde Q}^2$ and $M_{\tilde{\text{t}}}^2$ are the
{bilinear} soft supersymmetry-breaking terms for {$\tilde Q =
  (\tilde t, \tilde b)$ and $\tilde t_\text{R}$, respectively}.  We have
neglected \(D\)-term contributions, which are suppressed by gauge
couplings.

A convenient way to perform the summation is to solve
\begin{equation}\label{eq:tadpole}
G_1 =  -\text{i}\;\frac{\partial}{\partial h} V_1, \qquad
  \psset{unit=20mm}
  \begin{pspicture}(0,.45)(.6666,.6666)
     \psset{linewidth=.75pt, ArrowInside=->, arrowinset=0,linestyle=dashed}
     \psarc{>}(.5,.5){.16666}{-12}{180}
     \psarc(.5,.5){.16666}{180}{360}
     \psline(0,.5)(0.33333,.5)
  \end{pspicture}
\end{equation}
where $G_1$ is the Green function of the depicted $h$ tadpole with
field-dependent stop mass eigenstates propagating in the
loop, with $h$ defined after \eq{eq:PotentialFromGreensFunctions} \cite{Lee:1974fj}.
  Integrating \eqref{eq:tadpole} w.r.t.\ $h$ fixes the stop loop
  contribution $V_1^{{\tilde t}}$ to $V_1$ up to an arbitrary function
  of $h_\text{u}^0$. The correct dependence on $h_\text{u}^0$ is then
  found by deriving $V_1^{{\tilde t}}$ w.r.t.\ $h_\text{u}^0$ and
  $h_\text{u}^{0\dagger}$ and comparing the result with the
  $h_\text{u}^0$-$h_\text{u}^{0\dagger}$ two-point function depicted in
  the second row of \fig{fig:diagrams}.
 An alternative way to obtain the missing $h_\text{u}^0$-dependent piece, which leads to exactly the same result, uses the
replacement $h_\text{u}^0$ by $h_\text{u}^0 - w_\text{u}$. In the shifted theory this generates a three
point vertex $\tilde t_\text{L/R}^*$-$\tilde t_\text{L/R}$-$ h_\text{u}^0$ (and its
complex conjugate) generating in turn a tadpole diagram. The final result is found
after integration over $w_\text{u}$ and setting back $w_\text{u}=0$.
We find:
\begin{equation}\label{eq:result}
\begin{aligned}
V_1^{{\tilde t}} = &
 \frac{3{\widetilde{M}}_\text{t}^4}{32\pi^2} \bigg[
\left(1+x_\text{t}+y_\text{t}\right)^2
   \ln \left(1+x_\text{t}+y_\text{t}\right) \hfill\\[.5ex]
    &\qquad\; + \left(1-x_\text{t}+y_\text{t}\right)^2
    \ln\left(1-x_\text{t}+y_\text{t}\right)
\hfill\\[.5ex]
 & -\left(x_\text{t}^2+ y_\text{t}^2 + 2y_\text{t}\right)
 \left(3 - 2 \ln\left({\widetilde{M}}_\text{t}^2/Q^2\right)\right)
  \bigg],\hfill\\
\end{aligned}
\end{equation}
where the loops have been renormalized in the $\ov{\rm MS}/\ov{\rm DR}$ scheme
at the scale $Q$.  In
{\eq{eq:result}} we have used the mean soft mass square ${\widetilde{M}}_\text{t}^2\equiv
(M_{\tilde Q}^2+ M_{\tilde{\text{t}}}^2)/2$ and the dimensionless
quantities $x_\text{t}$ and $y_\text{t}$
\begin{equation}
\begin{aligned}\label{eq:abbrev}
x_\text{t}^2 & = \frac{\left|A_\text{t} h_\text{u}^0 {-} \mu^\ast Y_\text{t}
    {h_\text{d}^0}^\ast\right|^2}{{\widetilde{M}}_\text{t}^4} + \frac{(M_{\tilde Q}^2-
  M_{\tilde{\text{t}}}^2)^2}{4 {\widetilde{M}}_\text{t}^4} \hfill,\\[1ex]
 y_\text{t} & = \frac{\left|Y_\text{t}
h_\text{u}^0\right|^2}{{\widetilde{M}}_\text{t}^2}.
\end{aligned}
\end{equation}
An analogous expression (with obvious modifications) is found
  for the sbottom contribution $V_1^{{\tilde b}}$ and is given
    below. The shape of $V_1^{\tilde t}$
depends solely on the dimensionless parameters $x_\text{t}$ and $y_\text{t}$.  The
summation in \eqref{eq:PotentialFromGreensFunctions} converges if
${|x_\text{t}\pm y_\text{t}|} < 1$. The points $\pm x_\text{t}
  -y_\text{t}=1$ are branch points of the logarithm in the closed
result \eqref{eq:result}, which is the analytic continuation of the
  sum beyond the radius of convergence.
As we will argue below, the interplay between $V_0$ and $V_1$ can
lead to a potential with an unstable vacuum.

So far we have strictly argued along the line of deriving an
  effective Lagrangian and have resummed the higher-dimensional terms in
  $V_{D\geq 6}\subset{\cal L}_{\rm 2HDM}$, which arise from integrating
  out the heavy squarks. As long as one stays in this framework, one
  can deny any relevance of $V_1$ for large $h_{\text{u,d}}$ amplitudes with
  $|x_\text{t}-y_\text{t}|\geq 1$, because the series in
  \eq{eq:PotentialFromGreensFunctions} diverges in this domain.

The  justification of the use of $V_1$ for $|x_\text{t}-y_\text{t}|\geq 1$ lies
in the  effective potential definition of Coleman and Weinberg
  \cite{Coleman:1973jx}, which furthermore does not require the
  particles running in the loop to be heavy. We will later add the quark
  loops to $V_1$ to get the full one-loop effective potential $V_{\rm eff}$ in
  the sense of Coleman and Weinberg. {We briefly recall its derivation}.
Consider a theory with a complex scalar field
$\phi$. Connected Green functions can be derived by functional
  variations of a generating functional $W(J)$ w.r.t.\ a classical
  source $J(x)$.  The classical field $\phi_\text{c}$ is defined as the
expectation value of the field operator in the Fock vacuum in the
presence of the source $J$:
\begin{equation}\label{eq:phicl}
\phi _\text{c}  = \left[\frac{\langle \,0\,|\,\phi\,|\,0\, \rangle}{\langle
\,0\,|\,0\, \rangle}\right]_J.
\end{equation}
{A Legendre transform brings us to the \emph{effective action}
$\Gamma(\phi_\text{c}):= W(J)-\int d^4 x J(x) \phi_\text{c}(x)$, which is the generating
    functional of one-particle irreducible Green functions. The effective potential
    $V(\phi_\text{c})$ is defined as the first term of an
expansion of $\Gamma(\phi_\text{c})$ in terms of derivatives of $\phi_\text{c}$:
\begin{equation}
\Gamma(\phi_\text{c}) = \int d^4 x \lt[ -V(\phi_\text{c}) +
           \frac12 (\partial_\mu \phi_\text{c})^2 Z(\phi_\text{c}) +\ldots \rt]
\end{equation}
Thus the $n$-th derivative of $V(\phi_\text{c})$ is the sum of all one-particle irreducible graphs
with $n$ legs and zero external momenta.
$V(\phi_\text{c})$ can be
physically interpreted as follows \cite{Sher:1988mj}:}
The effective potential $V\left(\phi_\text{c}\right)$ is the potential
energy density of the classical field in the quantized theory,
\emph{viz.} the expectation value of the energy density in the state
$|\,0\,\rangle$ that minimizes $\langle\,0\,|\,H\,|\,0\,\rangle$ subject
to \eqref{eq:phicl} (where $H$ is the Hamiltonian density operator).
For vanishing sources $J\to 0$, the theory's vacua
seek to minimize the potential energy, \emph{i.e.}
\begin{equation}
  \label{eq:vacua}
        \frac{\delta V\left(\phi_\text{c}\right)}{\delta\phi_\text{c}} \,=\,0 .
\end{equation}
If this is the case for $\phi_\text{c} = \langle \phi\rangle \neq 0$,
the field takes a vacuum expectation value of $\langle \phi\rangle$, and
internal symmetries are broken spontaneously. If there is no asymmetric
vacuum in the classical potential, spontaneous symmetry breaking may
even emerge as a pure quantum effect.
The ground state {of the theory}, the
state of lowest energy, lives in the global minimum of the effective
potential \cite{Coleman-book}. Vacua minimizing the potential only
locally are unstable and can pass into the ground state.
Coleman and Weinberg~\cite{Coleman:1973jx} have considered the gauge
theory of a single scalar with self-interactions due to a classical
potential energy density $V_0(\phi)$.  They have found
\begin{equation}\label{eq:cwpotential}
   {V_{\rm eff}} \left( \phi_\text{c} \right) \,=\,
 \frac{1}{64\pi ^2} {\mbox{Tr}}\,
 \left[ V_0^{\prime\prime\, 2} ( \phi_\text{c} ) \,\ln
  V_0^{\prime\prime}( \phi_\text{c} )   \right] +
  P\left( \phi_\text{c} \right) ,
\end{equation}
where $P\left( \phi_\text{c} \right)$ is a polynomial depending on the
choice of the renormalization scheme.
$V_0^{\prime\prime}\left( \phi_\text{c} \right)$ is the field-dependent
mass matrix of the field degrees of freedom circulating in the loop,
{like the one in \eq{eq:fielddependentmass}.}
{To consistently include all terms involving $Y_\text{t}$ into
$V_{\rm eff}$ we must add the top loop,
$V_{\rm eff}=V_1^{\tilde t}+V_1^t$ to complement the result in
\eq{eq:result} to the full effective potential, with}
\begin{equation}\label{eq:top}
{
V_1^t = -\frac{3}{16\pi^2} \left|Y_\text{t} h^0_\text{u}\right|^4 \left[
\ln\left(\left|Y_\text{t} h^0_\text{u}\right|^2 / Q^2\right) {- \frac{3}{2}} \right]
}
\end{equation}
in the $\ov{\rm MS}/\ov{\rm DR}$ scheme. The generic formula in \eq{eq:cwpotential}
  (with the trace replaced by the supertrace to include the fermion
  loops) has been used in
  \cite{Brignole:1991pq} to derive $V_{\rm eff}$, expressed in terms of
  the eigenvalues of $\mathcal{M}^2_{\tilde{\mathrm{t}}}$ .  In our
  calculation, integrating the tadpole in \eq{eq:tadpole}, we find the same
  result. The MSSM case involves two Higgs fields $h$ and $h_\text{u}^0$ and thereby goes
  beyond the original framework in
  \cite{Coleman:1973jx}. Theories with several Higgs fields have been
  studied in \cite{Jackiw:1974cv}. Our situation with both 3-point and 4-point
  vertices present in the loops in \fig{fig:diagrams} is quite special
  and we have seen a benefit in verifying the commonly used formulae
for $V_{\rm eff}$ with our explicit calculation at the end of
Sec.~\ref{sec:el}: By deriving $V_1^{{\tilde t}}$ in \eq{eq:result}
w.r.t.\ $h$ and $h_\text{u}^{0\dagger}$ one indeed reproduces the coefficients
$a_{kn}$ in \eq{eq:akn} with the correct combinatorial factors.

In \eq{eq:result} we have given the explicit form of $V_1^{\tilde t}$
  in terms of $h^0_\text{u}$ and $h^{0*}_\text{d}$, which is more
  suitable for the global analysis of $V_{\rm eff}$. While we further
  generalize $V_1^{\tilde t}$ to complex $A_\text{t}$ and $\mu$, we neglect the
  small D-term contributions, which are included in
  \cite{Brignole:1991pq}.  Adding the contributions from (s)tops and
  (s)bottoms, the final ($\ov{\rm MS}/\ov{\rm DR}$-scheme) result reads
\begin{widetext}
\begin{equation}
\begin{aligned}
\label{eq:veff}
V_{\rm eff}  =\;& V_0+V_1^{\tilde t}+V_1^{t}+V_1^{\tilde b}+V_1^{b} \hfill \\[.5ex]
             =\;& m_{11}^{2\,^\text{tree}}\; |h_\text{d}^{0}|^2 +
  m_{22}^{2\,^\text{tree}}\; |h_\text{u}^{0}|^2 - 2 \operatorname{Re}\left(
    m_{12}^{2\,^\text{tree}} \; h_\text{u}^0 h_\text{d}^0 \right)
 +\frac{g^2+g^{\prime 2}}{8} \left( |h_\text{d}^{0}|^2 - |h_\text{u}^{0}|^2 \right)^2\hfill \\[.5ex]
 & + \frac{3{\widetilde{M}}_\text{t}^4}{32\pi^2} \bigg[
\left(1+x_\text{t}+y_\text{t}\right)^2
   \ln \left(1+x_\text{t}+y_\text{t}\right)
  + \left(1-x_\text{t}+y_\text{t}\right)^2 \ln \left(1-x_\text{t}+y_\text{t}\right) \hfill \\[.5ex]
  & \qquad\qquad\; - \left(x_\text{t}^2 + 2y_\text{t}\right) \left(
    3 - 2 \ln\left({\widetilde{M}}_\text{t}^2/Q^2\right) \right) -2 {y_\text{t}^2}
\ln\left(y_\text{t}\right)
\;+\;\{\text{t}\leftrightarrow\text{b}\}  \bigg]
\end{aligned}
\end{equation}
\end{widetext}
with $\widetilde{M}_{\text{t},\text{b}}^2 = (M_{\tilde Q}^2 +
M_{\tilde{\text{t}},\tilde{\text{b}}}^2)/2$ and stop-loop parameters $x_\text{t}$ and $y_\text{t}$ defined as in (\ref{eq:abbrev}); similarly, the sbottom-loop parameters are
\begin{equation}
\begin{aligned}
x_\text{b}^2 & = \frac{\left|A_\text{b} h_\text{d}^0 {-} \mu^\ast Y_\text{b}
    {h_\text{u}^0}^\ast\right|^2}{{\widetilde{M}}_\text{b}^4} + \frac{(M_{\tilde Q}^2-
  M_{\tilde{\text{b}}}^2)^2}{4 {\widetilde{M}}_\text{b}^4} ,\hfill \\[.5ex]
y_\text{b} &= \frac{\left|Y_\text{b}
h_\text{d}^0\right|^2}{{\widetilde{M}}_\text{b}^2}.
\end{aligned}
\end{equation}
In \eq{eq:veff}, the quadrilinear couplings at tree-level
\(\lambda_i^\text{tree}\) were replaced
according to \eq{eq:MSSM-rep} and the treatment of the bilinear
\(m_{ij}^{2\,^\text{tree}}\) is discussed in the following section. Note,
that the \(\lambda_4^\text{tree}\) term drops out from the neutral Higgs
potential.

Apparently, the effective potential in \eq{eq:cwpotential} acquires an
imaginary part for values of $\phi_\text{c}$ which render an eigenvalue of
$V_0^{\prime\prime\, 2} ( \phi_\text{c})$ negative. In our case of $V_1^{\tilde
  t}$ this happens for $x-y>1$. The imaginary part of $V_{\rm eff}$
  must be dropped to keep the Lagrangian hermitian.  The physical
  meaning of the imaginary part is controversial: Weinberg and Wu
consider the case of a theory with a single scalar field $\phi$ and
argue that the imaginary part of $V_{\rm eff} (\phi_\text{c})$ coincides with
the decay rate of a particular quantum state $\ket{\eta}$ satisfying
$\bra{\eta} \phi \ket{\eta}=\phi_\text{c}$ \cite{Weinberg:1987vp}.  A
  different viewpoint on the imaginary part is expressed in
  \cite{patel}. We remark that (the principal values of) analytical
  continuations are ambiguous, \emph{e.g.}\ replacing $\ln(1-x_\text{t}+y_\text{t})$ by $1/2
  \ln(1-x_\text{t}+y_\text{t})^2$ in \eq{eq:result} does not change the series
  expansion in \eq{eq:PotentialFromGreensFunctions}, but relocates the
  branch cut in a way that $V_{\rm eff}$ stays real for
  $|1-x_\text{t}+y_\text{t}|>1$.

\section{Phenomenology of the MSSM vacuum instability\label{sec:ph}}
In this section we give explicit examples for MSSM parameters
 leading to a $V_{\rm eff}$ for which ``our'' vacuum with
$v=246\gev$ is unstable.

\begin{figure*}[t]
  \centering
\includegraphics{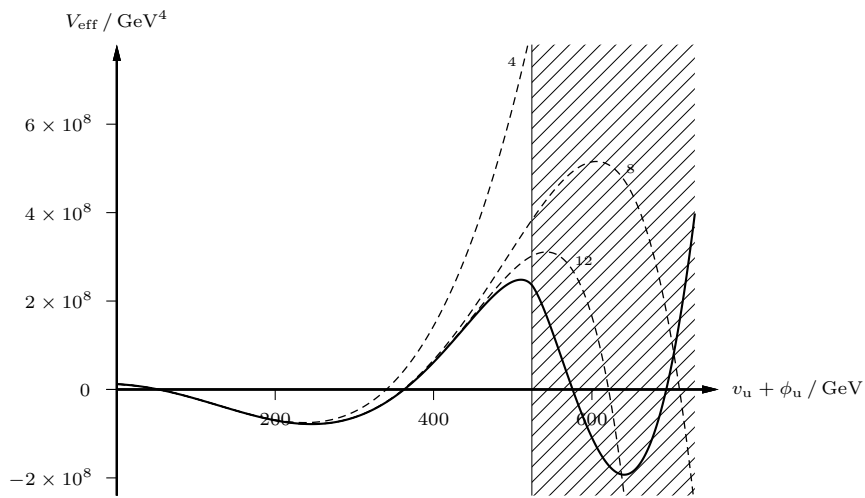}
  \caption{The 1-loop effective potential $V_\text{eff}= V_0+V_1$ for
    $\tan\beta = 40$ and $m_{A^0} = 800 \,\text{GeV}$.
    Soft supersymmetry-breaking masses as well as the renormalization scale $Q$ have
been taken at
    $1\,\text{TeV}$. The Higgs couplings in the loops with top and
    bottom squarks involve
    $\mu = 2.55\,\text{TeV} $ and $A_\text{t} \simeq 1.5
    \,\text{TeV} $. The hatched area highlights the analytic continuation
    beyond the branch point at $x-y=1$.  The cases with
    {$V_{\rm eff}$} truncated after terms of dimension
    {$2(k+n)=4,8,$ and $12$ are shown with dashed lines}
    (from top to bottom).}\label{fig:potentialA}
\end{figure*}

{While loop corrections can render the parameters in {\eq{eq:V2HDM}}
complex \cite{Ellis:2007kb,Gorbahn:2009pp}, we restrict ourselves to
the case of real parameters, with a}
mass matrix that does not mix \CP~eigenstates.
{Writing}
\begin{equation}\label{eq:FluctuationFields}
\begin{aligned}
 h_\text{u}^0 & = \frac{1}{\sqrt{2}} \left(v_\text{u} + \phi_\text{u} +
 \text{i}\chi_\text{u}\right), \hfill\\[.5ex]
  h_\text{d}^0 & =
 \frac{1}{\sqrt{2}} \left(v_\text{d} + \phi_\text{d} +
 \text{i}\chi_\text{d}\right),
\end{aligned}
\end{equation}
we trade two of the mass parameters in \eq{eq:veff}
for $v_{u,d}$ in analogy to \eq{eq:m11vev}:
\begin{widetext}
\begin{equation}\label{eq:min-cond}
\begin{aligned}
 m^{2\,^\text{tree}}_{11} &= m^{2\,^\text{tree}}_{12}\,\tan\beta -\frac{v^2}{2}
 \cos(2\beta) \lambda^{\text{tree}}_1 - \frac{1}{v\cos\beta}\,\left.\frac{\delta}{\delta\phi_\text{d}}V_1\right|_{\substack{ {\phi_{\text{u},\,\text{d}} \,\to\, 0}\\ {\chi_{\text{u},\,\text{d}}  \,\to\, 0}}},\hfill\\
 m^{2\,^\text{tree}}_{22} &= m^{2\,^\text{tree}}_{12}\,\cot\beta +\frac{v^2}{2}
 \cos(2\beta) \lambda^{\text{tree}}_1 - \frac{1}{v\sin\beta}\,\left.\frac{\delta}{\delta\phi_\text{u}}V_1\right|_{\substack{ {\phi_{\text{u},\,\text{d}} \,\to\, 0}\\ {\chi_{\text{u},\,\text{d}}  \,\to\, 0}}}.\hfill\\
\end{aligned}
\end{equation}
\end{widetext}
The Higgs mass matrices are
\begin{equation}
\begin{aligned}\label{eq:mm}
M^2_{\text{R}\,ij} &= \left.\frac{\delta^2 V}{\delta\phi_i\,\delta
    \phi_j}\right|_{\substack{ {\phi_{\text{u},\,\text{d}} \,\to\, 0}\\ {\chi_{\text{u},\,\text{d}}  \,\to\, 0}}}, \hfill\\[.5ex]
M^2_{\text{I}\,ij} &= \left.\frac{\delta^2 V}{\delta\chi_i\,\delta
    \chi_j}\right|_{\substack{ {\phi_{\text{u},\,\text{d}} \,\to\, 0}\\ {\chi_{\text{u},\,\text{d}}  \,\to\, 0}}}.
\end{aligned}
\end{equation}
The mass of the pseudoscalar Higgs $A^0$ is essentially
controlled by $m_{12}^{2\,^\text{tree}}$ (for not too small $\tan\beta$). The mass
of the lightest scalar Higgs $h^0$ is very sensitive to radiative
corrections, and a calculation from the one-loop effective potential
  through \eq{eq:mm} underestimates $m_{h^0}$ by more than 5$\gev$.
We, accordingly, can safely calculate \(m_{A^0}\) from the non-zero
eigenvalue of $M^2_{\text{I}}$, but resort to \texttt{FeynHiggs
  2.10.0}
\cite{Heinemeyer:1998np,Heinemeyer:1998yj,Degrassi:2002fi,Frank:2006yh}
to obtain \(m_{h^0}\) including dominant 2- and 3-loop
contributions. While the smallness
   of gauge couplings governing the tree-level result for $m_{h^0}$
   makes the need of higher-order corrections obvious, one does not
   expect large radiative corrections to the height of the second minimum,
   which involves large parameters ($\mu$, $A_\text{t}$, $m_{\tilde t_{\text{L},\,\text{R}}}^2$) already at
the leading one-loop order. The different mass scales
   entering our analysis are close enough that no large logarithms occur
   making any renormalization-group improvement obsolete. The scale $Q$
   entering $V_{\rm eff}$ explicitly and the couplings and mass parameters
   implicitly is taken at $\widetilde{M_{\text{t}}}$.
{We are interested in a heavy $A^0$ (to satisfy the experimental
  constraints from $A^0\to \tau\tau$  and $B_s\to \mu^+\mu^-$)
in which case the effect of $m_{A^0}$} on the lightest Higgs
mass is small. After choosing values for $\mu,M^2_{\tilde
    Q,\tilde{\text{t}},\tilde{\text{b}}},m_{A^0}$, and $\tan\beta$ we adjust
$A_\text{t}$ to fit $m_{h^0}=
126\,\text{GeV}$. Finally we remark that we take the gluino mass heavier
than the squark masses. The gluino mass enters the threshold corrections
subsumed in the parameter $\Delta_\text{b}$ which appears in the bottom Yukawa
coupling as $Y_\text{b}=m_\text{b}/(v_\text{d}(1+\Delta_\text{b}))$
\cite{Hall:1993gn,Carena:1994bv,cgnw,Hofer:2009xb} and for sufficiently
large gluino mass we can neglect $\Delta_\text{b}$.

\begin{figure*}[t]
  \centering
\includegraphics{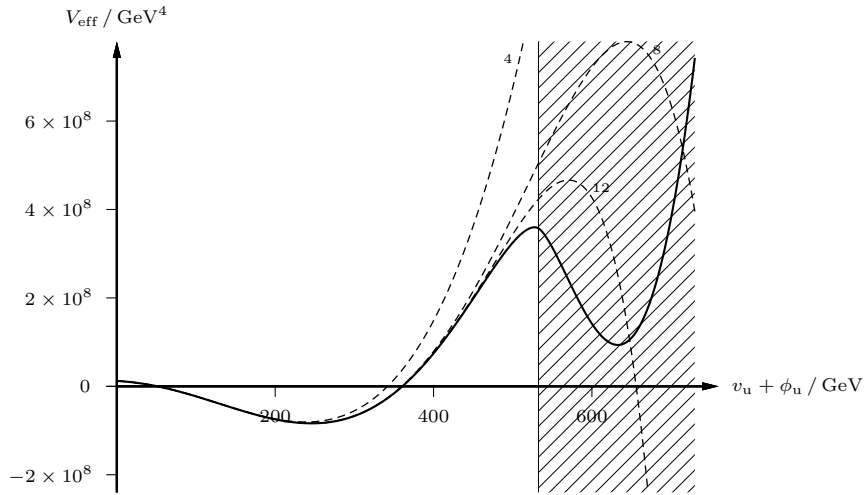}
  \caption{The 1-loop effective potential as in \fig{fig:potentialA},
  with $\mu$ lowered to $2.51\,\text{TeV} $.}
\label{fig:potentialB}
\end{figure*}

The 1-loop effective scalar potential obtained with this setup is
illustrated in Figs.~\ref{fig:potentialA} and \ref{fig:potentialB}
 for a sample
  parameter point complying with all experimental constraints. We find
that supersymmetric quantum effects lead to a second, deeper minimum of
the Higgs potential.
Relative depth and position of the minima
depend crucially on the values of $\mu,A_{\text{t,b}},$ and $\tan\beta$.
Large values parametrically enhance the effect of the sfermion loops.

Note that the minimum of the 1-loop contribution is always beyond the
branch point of one of the logarithms in $V_{\rm eff}$.
For the case depicted in \fig{fig:potentialA} the electroweak
vacuum is unstable and a transition to the ground state is possible due
to quantum tunnelling
\cite{Kobzarev:1974cp,Coleman:1977py,Coleman:1977py-erratum,Callan:1977pt}.
Semiclassical methods can be applied to estimate the lifetime of the
unstable vacuum.  If the inferred lifetime exceeds the age of the
universe, the instability of the electroweak vacuum does not
  matter. However, in the case at hand, with the global minimum
  appearing for field values of less than $700\gev$, the electroweak
  vacuum is extremely short-lived.
\fig{fig:potentialA} further shows that the
loop-corrected 2HDM potential in \eq{eq:V2HDM} (corresponding to a
truncation of  $V_{\rm eff}$ at terms of dimension 4) does not
reveal any problems. It is therefore not sufficient to include
loop corrections to the 2HDM parameters to analyze the vacuum stability.
Including a finite number of higher-dimension terms leads to a Higgs
potential which is unbounded from below (UFB), but the correct feature of a
potential with a second minimum is only found from the full $V_{\rm
  eff}$. This is not surprising, since the second minimum is in the
domain beyond the radius of convergence of the series in
\eq{eq:PotentialFromGreensFunctions}.
Lowering $\mu$ slightly raises the second minimum and leads to {a}
parameter point passing our criterion, as depicted in \fig{fig:potentialB}.
Note that here the UFB criterion applied to a Higgs potential
truncated at a finite dimension $2(k+n)$ leads to a premature
exclusion of the corresponding MSSM parameter point.

In the region beyond the {branch} point, {$V_{\rm eff}$ in
\eq{eq:veff}} features an imaginary part, which should be dropped from
the Lagrangian. Here the field-dependent
squark mass matrix of \eqref{eq:fielddependentmass}---which is the second
derivative of the scalar potential
with respect to the sfermion fields---acquires a negative
eigenvalue. If we depicted the sfermion field corresponding to
this tachyonic mass eigenstate perpendicular to the drawing plane
in \fig{fig:potentialA}, the minimum in the $\phi_\text{u}$ direction
is revealed as a local maximum in the sfermion direction (\emph{i.e.}\ we
encounter a saddle point of the full scalar potential) and the global
minimum of the scalar potential will be necessarily a charge and color breaking (CCB) vacuum.
 The example of \fig{fig:potentialB} also shows that the
  existence of an imaginary part in {$V_{\rm eff}$} alone
  does not directly lead to an unstable
  vacuum.

There exist several criteria in the literature to check whether or not
the parameters lead to a CCB vacuum at tree-level. We can easily check,
that we are in full agreement with the traditional criterion
\cite{Gamberini:1989jw,Casas:1995pd} \(A_\text{t}^2 < 3 (M^2_{\tilde Q}
+ M^2_{\tilde{\text{t}}} + m_{22}^{2\,^\text{tree}})\).
A stronger empirical bound of \(A_\text{t}^2 + 3\mu^2 \lesssim 7.5
  (M^2_{\tilde Q} + M^2_{\tilde{\text{t}}})\), which our sample point
  would not pass, has been suggested in \cite{Kusenko:1996jn}.  However,
  this bound has been critically reviewed in the recent detailed
  analysis \cite{Blinov:2013fta}, which advocates bounds closer to the
traditional measure. The criterion of \cite{Blinov:2013fta}
  translates to \(A_\text{t}^2 \lesssim 3.4 (M^2_{\tilde Q} +
  M^2_{\tilde{\text{t}}}) + 60 m_{22}^{2\,^\text{tree}}\) in our case
  and is fulfilled by the parameters of Figs.~\ref{fig:potentialA} and
\ref{fig:potentialB}. We are therefore safe from CCB minima
  of the tree-level potential.

\begin{figure*}[tb]
  \centering
\includegraphics[width=0.7\textwidth]{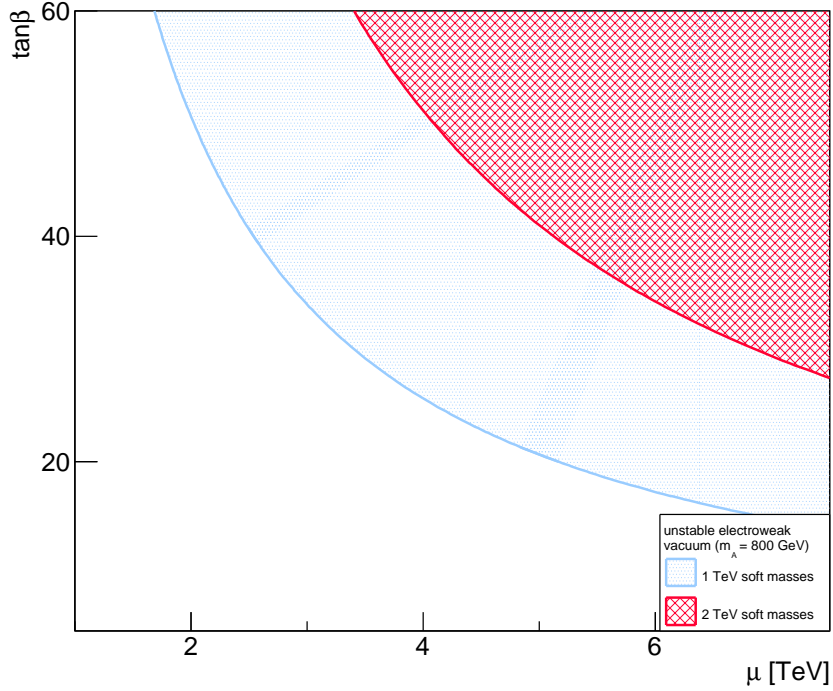}
  \caption{{Area in the $\mu$--$\tan\beta$ plane for which
$V_{\rm eff}$ develops an unwanted minimum as depicted in
      \fig{fig:potentialA}. The red, cross-hatched area corresponds to
$M_{\tilde Q}=M_{\tilde{\text{t}}}=M_{\tilde{\text{b}}}{=Q}=2\tev$; the light blue area
is excluded if $M_{\tilde Q,\tilde{\text{t}},\tilde{\text{b}}}$ is lowered to 1$\tev$.
$A_\text{t} \simeq 1.5
    \,\text{TeV} $ is fitted to reproduce $m_{h^0}= 126\gev$ in both cases;
$m_{A^0}=800\gev$ is chosen to comply with LHC search limits for
$A^0\to\tau\tau$.}}
\label{fig:excl}
\end{figure*}

Whenever the situation depicted in \fig{fig:potentialA} occurs the
  corresponding MSSM parameter point is excluded. We show the excluded
  region of the $\mu$--$\tan\beta$ plane in \fig{fig:excl} for two
  values of the squark masses.  We stress that the consideration of a
  single direction in the multi-dimensional space of scalar fields is
  not sufficient to prove the stability of the electroweak
  vacuum. \emph{I.e.}\ to validate or discard the MSSM parameter point of
  \fig{fig:potentialB} one would have to study all directions in the
  $h_\text{u}^0$--$h_\text{d}^0$ plane. A complete investigation further requires the
  study of the global minimum of the full scalar potential (including
  the sfermion fields) with the field-dependent sfermion masses (see
  \eq{eq:fielddependentmass}): As discussed above in conjunction with
  the second minimum of $V_{\rm eff}$, the sfermion potential is
  non-convex in the region with large Higgs fields with the possibility
  of a CCB  minimum below the desired ground state of the electroweak
  vacuum. The determination of the global minimum of the loop-corrected
  full scalar potential is a formidable task and
  beyond the scope of this paper.
An accurate determination of the
  contours delimiting the allowed parameter space in \fig{fig:excl}
  may also require to use the renormalization-group improved two-loop result for
$V_{\rm eff}$ \cite{Martin:2002i}.

The requirement of a stable vacuum excludes large values of
  $\mu\tan\beta$. The sample points studied by us also involve a large
  value of $A_\text{t}$, to accommodate $m_{h^0}= 126\gev$ through sizable stop
  mixing. This portion of the MSSM parameter space is of interest in
  flavor physics and has been widely studied: The product
  $A_\text{t}\mu\tan\beta$ governs the size of the chargino contributions to
  $B(B\to X_s \gamma)$ \cite{Ciuchini:1998xy,Degrassi:2000qf,cgnw2} and
  the Higgs-mediated contributions to $B(B_{d,s}\to \mu^+\mu^-)$ and
  \bbms\ \cite{Babu:1999hn,bcrs,Isidori:2001fv,Buras:2002wq,
    Dedes:2002er,brs,Hofer:2009xb,Gorbahn:2009pp,Crivellin:2010er} grow
  with $A_\text{t}$, $\mu$ and higher powers of $\tan\beta$ (see \cite{Altmannshofer:2013oia} for a recent
  study).   Similar to the quark sector,
  flavor-changing neutral current processes in the lepton sector can be
  enhanced if $\mu \tan\beta$ is large
  \cite{Masina:2002mv,Paradisi:2005fk,Girrbach:2009uy}.  Therefore the
  global minimum of $V_{\rm eff}$ should be checked in MSSM parameter
  scans of flavor observables.

\section{Conclusions}
The MSSM Higgs potential receives large radiative corrections from loops
with stops and (if $\tan\beta$ is large) sbottoms.  Squarks which are
much heavier than the Higgs bosons can be integrated out resulting in an
effective Lagrangian of a two-Higgs-doublet model. The Lagrangian can be
systematically improved by higher-dimensional terms suppressed by powers
of $1/M_{\tilde Q, \tilde{\text{t}}}^2$. We have calculated the stop contribution to
the effective self-couplings $(h_\text{u}^{0\dagger} h_\text{u}^0)^{d_1}
(h_\text{d}^{0\dagger} h_\text{d}^0)^{d_2}$ of any number of neutral $h_\text{u}^0$ or
$h_\text{d}^0$ fields at the one-loop level using an elementary diagrammatic
method.  Depending on the MSSM parameters entering the loop diagrams,
the Higgs potential of the resulting effective Lagrangian can be
unbounded from below or feature a second, unwanted minimum which is
deeper than the one with $\sqrt{|h_\text{u}^{0}|^2+|h_\text{d}^{0}|^2}=v=246\gev$.
In this paper we have found that one cannot assess the question of
vacuum stability from such an effective Lagrangian truncated at a finite
dimension $2(d_1+d_2)$: the critical values of $h_\text{u,d}^{0}$ for
which the Higgs potential drops below its value at $v=246\gev$ are
beyond the radius of convergence of the sum over $d_1$, $d_2$.

The effective potential $V_{\rm eff}$ sums the squark-induced
higher-dimensional Higgs self-couplings to all orders (without the need
of any hierarchy between squark and Higgs masses) and permits the proper
inclusion of top loops as well. We find that the one-loop MSSM effective
potential is bounded from below but develops a second, deeper minimum,
if the parameters $\mu$, $\tan\beta$, or $A_\text{t}$ governing the
squark-Higgs couplings become too large (see \fig{fig:potentialA}).  For
two values of degenerate squark masses we have determined the region in
the $\mu$--$\tan \beta$ plane corresponding to an unstable vacuum (see
\fig{fig:excl}). $A_\text{t}$ has been chosen to reproduce the correct mass of
126\gev\ for the lightest neutral Higgs boson, which drives $|A_\text{t}|$ to
large values. The excluded region of large $|\mu|\tan\beta$ and large
$|A_\text{t}|$ is widely studied in flavor physics, since in this region the
MSSM contributions to several flavor-changing processes are large. We
argue that the criterion of a global minimum of $V_{\rm eff}$ with
$v=246\gev$ should be included in phenomenological analyses determining
the allowed parameter space of the MSSM.

\section*{Acknowledgements}
We are grateful for discussions on various aspects of the
effective potential with L.~Di Luzio, L.~Mihaila, S.~Pokorski, and
M.~Zoller. M.B.\ and W.G.H.\ acknowledge financial support of
\emph{Studienstiftung des deutschen Volkes}\ and the DFG-funded
Research Training Group GRK 1694. G.C thanks W. Porod for useful discussions and
clarifications about the {\tt Vevacious} code. This work was supported by BMBF under
grant no.~05H12VKF {and the Theory-LHC-France initiative of the CNRS/IN2P3.}

\bibliography{literature}

\end{document}